\documentclass{elsart}

\usepackage{graphicx, amsmath,amssymb, mcite}
\usepackage{lineno}
\linenumbers
\begin{document}

\begin{frontmatter}

\title{Periodical cicadas: a minimal automaton model}

\author[IFSC]{Giovano de O. Cardozo}
\author[IFSC]{Daniel de A. M. M. Silvestre\corauthref{CA}}
\author[UEFS]{Alexandre Colato}

\corauth[CA]{Corresponding author: silvestre@ifsc.usp.br}

\address[IFSC]{Instituto de F\'isica de S\~ao Carlos, Universidade de S\~ao Paulo, Caixa Postal 369, 13560-970 S\~ao Carlos SP, Brazil}
\address[UEFS]{Universidade Estadual de Feira de Santana, Departamento de F\'isica. Campus Universit\'ario, BR 116, KM 03, 44031-460, Feira de Santana BA, Brazil}

\begin{abstract}
The \textit{Magicicada spp.} life cycles with its prime periods and highly synchronized emergence have defied reasonable scientific explanation since its discovery. During the last decade several models and explanations for this phenomenon appeared in the literature along with a great deal of discussion. Despite this considerable effort, there is no final conclusion about this long standing biological problem. Here, we construct a minimal automaton model without predation/parasitism which reproduces some of these aspects. Our results point towards competition between different strains with limited dispersal threshold as the main factor leading to the emergence of prime numbered life cycles.
\end{abstract}

\begin{keyword}
\textit{Magicicada} \sep celular automaton \sep patch dynamics \sep competition
\PACS 87.10.+e \sep 87.23.n \sep 07.05.Tp
\end{keyword}

\end{frontmatter}

\section{Introduction}
\label{intro}
The origin and evolution of the \textit{Magicicada spp.} life cycles is one of the most intriguing problems in population biology and evolution. These long term periodical life cycles with prime period (namely 13 and 17 years) and the incredibly synchronized emergence of the adults have defied all attempts of ultimate explanation since their discovery some 300 years ago \cite{Ziebarth2005}. During the last 15 years a plethora of models and possible explanations for this phenomenom appeared in the literature (e.g. \cite{Ziebarth2005, Goles2000, Behncke2000, Webb2001, Hayes2004, Campos2004} and \cite{Williams1995, Heliovaara1994} for a good review). However, despite this considerable effort, there is no final conclusion about this long standing biological problem. Currently, there seems to be two main lines debating this subject. The traditional line advocates that this type of life cycle emerges as response of the cicadas against predation pressure and limited resources \cite{Behncke2000, Campos2004, Lloyd1966a, Lloyd1966b, Hoppensteadt1976, Bulmer1977}. Thus, a prime numbered life cycle with highly synchronized emergence is thought to be a strategy to evade predation by minimizing the probability of interspecific interaction and promote predator satiation during population exposure at the adult part of its life cycle. On the other hand, some authors propose that this type of life history emerges to avoid hybridization between the different strains of cicadas under harsh environmental conditions\cite{Cox1988, Yoshimura1997, Cox1998}. Specifically, environmental conditions had led to delayed emergence and limited mating opportunities during ice age periods and this promoted synchronization in populations with periodic life cycles. In this scenario, the prevalence of prime numbered life cycles is explained by their low probability of hybridization with other life cycles. It's important to state that, by definition, an insect is said to be periodic if its life cycle has a fixed length of $k$ years ($k>1$) and adults do not appear every year but only every $k$th year. Otherwise, we call that insect annual, despite of the length of its life cycle (cf. \cite{Bulmer1977}).

Recently, three accounts on the subject were published \cite{Ziebarth2005, Campos2004, Grant2005} suggesting a somewhat different line of thought. Those authors believe that competition is the main factor leading to periodicity as defined above, based on the assumption that competition between different strains is stronger than competition within a specific strain \cite{Bulmer1977}. They suggest competition between strains with nymphs of other cicada species (outside the \textit{Magicicada} group) would enhance selection for periodicity by augmenting the intensity of intraspecific competition and determining the spatial distribution of the strains. The emergence of prime periods would either be just an artifact of the process \cite{Ziebarth2005} or even does not need an explanation at all \cite{Grant2005}. In \cite{Campos2004}, the model used deal with most aspects reviewed here in a very simple and clear manner. One can verify that the assumptions made by those authors are, indeed, biologically reasonable. Nevertheless, the problem still persists. What are the sufficient conditions for the emergence of prime numbered life cycles? Which mechanisms are responsible for that? To what extent? In this contribution, we will try to address some of these questions in a straightforward manner.

%%%%%%%%%%%%%%%%%%%%%%%%%%%%%%%%%%%%%%%%%%%%

\section{The Model}
\label{model}
Our model is inspired on the works of Campos \textit{et al.} and Goles \textit{et al.} \cite{Goles2000, Campos2004} with some simplifications and a rather different biological interpretation. Instead of a individual-based population dynamics, our model consists of very simple patch dynamics in the spirit found in \cite{Burrows1998}. Based on \cite{Grant2005}, we assume competition as the principal ingredient in this scenario. In this way, the dynamics presented here do not include any type of antagonistic interaction besides the competition between the strains. Therefore, we construct a stochastic cellular automaton with periodic boundaries on a squared lattice of linear dimension $L$. Each lattice site represents one habitat patch. At a given generation, a patch may be empty ($s_{i}(t)=0$) or colonized ($s_{i}(t)=1$) by a subpopulation. If this is the case, the colonized patch has two more characteristics: a life cycle $k=2\ldots,d$ defined by its length in generations and an age $t_i(t)$. The parameter $d$ stands for the total diversity of life cycles. The update of each patch runs in parallel and each generation (our discrete time step) consists of a complete lattice update. At each generation step, all occupied sites have its age incremented by $1$. When colonized patch has age equal to its life cycle length ($t_i(t)=k$) we say that it is in the active state. Biologically, this corresponds to the adult part of the cicada life cycle. Individuals can only interact directly during this phase of its life cycle. On the other hand, every time an empty patch (innactive site) is found we look at its closest neighborhood (Moore neighborhood with range $1$) and count the number of active patches. If the number of these is greater than the dispersal threshold parameter $M$, that empty patch will be eventually colonized. After this, a randomly chosen active patch is picked from the neighborhood and that individual will be responsible for the colonization of the empty patch. The newly colonized patch has the same life cycle length of its colonizer and age set to zero. This process is biologically reasonable and mimics very well a competitive dynamics between the different strains of cicadas. The parameter $M$ can be viewed as measure of a tendency for dispersal of the population. Therefore, for small $M$ there is a high tendency for dispersal and we need small populational density to have that. Conversely, a large $M$ implies in a high populational density in order to promove dispersal. At the end of a generation step, all active patches have their age set to zero and the whole process begin again.

For each simulation run, a fraction $x_{0}$ of the lattice is initially occupied, randomly. For each of occupied patch, a life cycle and an age are selected, in this order, from a discrete uniform distribution according to the limits imposed by the parameter $d$. Therefore, the initial population is a random mixture of all possible life cycles in a complete desynchronized fashion. Our main interest is to study the long term behavior of this kind of system and to verify whether we can recover the results found in \cite{Goles2000, Campos2004} in this simplified scenario. Henceforth, for each generation step we count the life cycles present in the population. The life cycle which makes up the largest fraction of the lattice at that generation step is the winner at that time, i.e., a local winner. We proceed this way until the winning life cycle stops changing, thus, becomes the global winner. Of course, if two even life cycles (e.g., $k=2$ and $k=4$) have an odd emergence phase shift, they will never encounter each other and are completely unable to compete directly. This situation never happens between two prime numbered life cycles.

%%%%%%%%%%%%%%%%%%%%%%%%%%%%%%%%%%%%%%%%%%%%

\section{Results and Discussion}
\label{results}
For our simulation runs we used a maximum generation time ($t_{max}$) of $10^{6}$, which proved to be enough simulation time to find a global winner (data not shown). We set $L=100$ and $d=24$ for all simulation runs performed. The other parameters were varied to observe the effects of different initial population size and dispersal threshold. It's important to point out that for each run a parameter set is kept fixed. For each parameter set $1000$ independent runs were executed.

First of all, let's explore the effect of different initially occupied fraction of the lattice at fixed dispersal threshold $M$. We can observe in Fig. 1 the very sharp rising of the occupied fraction $x$, starting the simulation with $x_{0}=0.1$ and a much slower variation in the case $x_{0}=0.5$. This difference is explained easily when one looks at global winner distribution of both situations. Starting with a small $x_{0}$, the rapid spread of short lived strains is facilitated. However, this spreading is clearly cooperative as suggested by the sharp rising curve. An increasing in the short lived strains implies in a greater probability of colonization and vice-versa. On the other hand, a larger $x_{0}$ geometrically prevents this fast spreading simply because the clusters of short lived strains are now blocked by clusters of long lived strains. Of course, even in this condition, short lived strains are commonly the global winners. But now, we can see a more varied distribution of winners. Compare Fig. 2 and Fig. 3.

In second place, we start to observe the effect of varied dispersal threshold. A clear predominance of short lived cicadas as the global winners is seen for $M=2$. It could not be different. A small dispersal threshold requires low populational densities, as said before, to ensure colonization. Consequently, life cycles more active on average (i.e., the short ones) tend to spread rapidly over the empty patches before any reaction from the other life cycles. This is exactly what is observed in Fig. 4. Setting $M=3$ changes completely the scenario. In Fig. 5, one can see an evident hegemony of prime numbered life cycles. With this parameter set, on average, each active patch will compete with more than three other active sites for colonization. Therefore, competition is in a much higher level than in the $M=2$ case. Now, let's turn our attention to the $M=4$ case. As seen before, there is a predominance of cicadas with prime numbered life cycles as the global winners. Moreover, the majority of life cycles are well represented in the global winners histogram (Fig. 6). It is important to note that at this level of dispersal threshold is virtually impossible to fill up the entire lattice. In fact, the initial population grows just marginally before reaching the steady state. This is due to a geometrical border effect. In such case, the growth of the global population is strongly self-limited. The same will occur to $M>4$. Actually, for $M>4$ no appreciable growth and/or spread of the population could be observed.

\section{Conclusion}
\label{conclusion}

In the present contribution, we showed that a very simple competitive dynamics, spatially structured, with few parameters can exhibits a reasonable diversity of behaviors. But, the main point here is that, diferently from the majority of works on this subject, we  demonstrated in a simple and direct manner the insufficiency of predation to ensure the emergence of prime numbered life cycles as the most effective ones in the dynamic. In our model, in which only competition can change the fate of the different strains, the simplest way of avoiding competition is to reduce the chance of interaction between different strains. For this purpose, prime numbered life cycles have the least tendency for interaction in the long run. And more, this model indicates that prime numbered life cycles experience a type of kin selection. In this manner, they tend to interact preferentially with other prime numbered ones rather than with non prime numbered. In the end, there's no need for \textit{ad hoc} explanations for the success of those life cycles. Our result points towards competition between the different strains as responsible for the emergence of prime numbered life cycles. This results contrasts sharply with those in \cite{Goles2000, Campos2004}, in which a much more complicated dynamics is explored. Specifically, we reproduced the results of \cite{Campos2004} with and without the presence of predators. The only detectable difference was a shift to the right in the global winner histogram (data not shown here). It could not be different, as the chance of interaction is high between short life cycle strains and predators. In this respect, our model could be seen as a reinterpretation of the models presented in \cite{Goles2000, Campos2004} without mutation and predation. But, as one can see, we obtained very similar results. Finally, we hope that this simple contribution can help to elucidate this very interesting puzzle of Nature by showing how simple mechanisms can generate unexpected (and amazing) results.

\section*{Acknowledgements}
We thank P. R. A. Campos for valuable discussions and for initial inspiration. The work of D A. M. M. S. is supported by CAPES. G. O. C. and A. C. were supported by FAPESP.
\label{acknow}

\newpage

  Fig. 1: Temporal evolution of the occupied fraction of the lattice ($x$) for dispersal threshold $M=2$ and initial occupied fraction $x_{0}$ as indicated in the graph. This graph is for just one run, but it represents significantly the model's general behavior.\\
\newline
  Fig. 2: Distribution of the global winner for initial occupied fraction $x_{0}=0.1$ and dispersal threshold $M=2$ in $1000$ independent runs.\\
\newline
  Fig. 3: Distribution of the global winner for initial occupied fraction $x_{0}=0.5$ and dispersal threshold $M=2$ in $1000$ independent runs.\\
\newline
  Fig. 4: Distribution of the global winner for initial occupied fraction $x_{0}=0.5$ and dispersal threshold $M=3$ in $1000$ independent runs. The dominance of prime numbered life cycles is evident.\\
\newline
  Fig. 5: Distribution of the global winner for initial occupied fraction $x_{0}=0.5$ and dispersal threshold $M=4$ in $1000$ independent runs. Again, the dominance of prime numbered life cycles is evident, but, to a lesser extent in this case.\\
\newline
  Fig. 6: Temporal evolution of the occupied fraction of the lattice ($x$) for initial occupied fraction $x_{0}=0.5$ and dispersal threshold as indicated in the graph. As stated in sec. \ref{results}, the growth for $M=4$ is strongly limited.\\
\newline
\newpage
\begin{center}
  \begin{figure}
    \resizebox{0.75\columnwidth}{!}{%
      \includegraphics{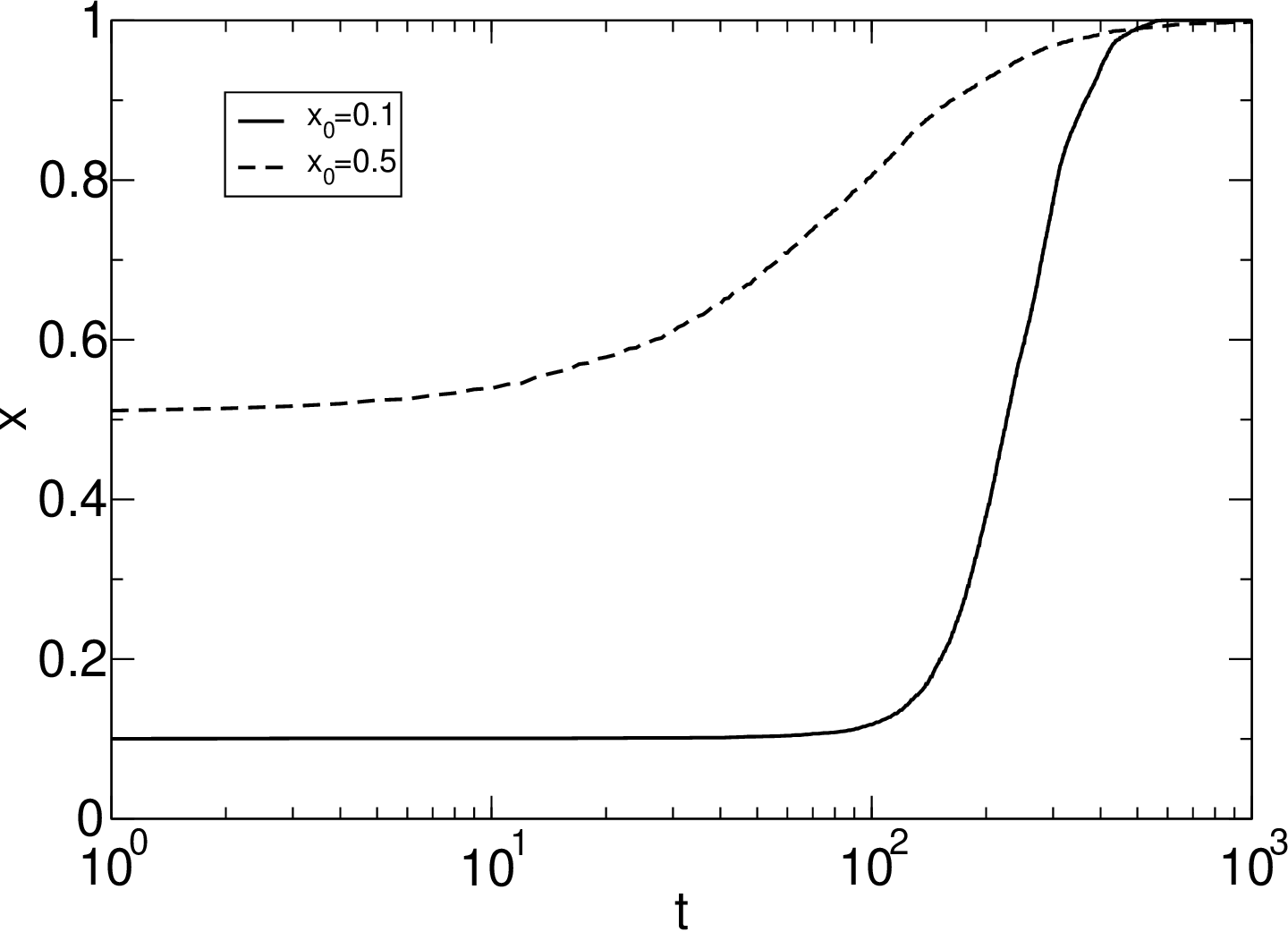}
    }
  \end{figure}
\end{center}

\newpage
\begin{center}
  \begin{figure}
    \resizebox{0.75\columnwidth}{!}{%
      \includegraphics{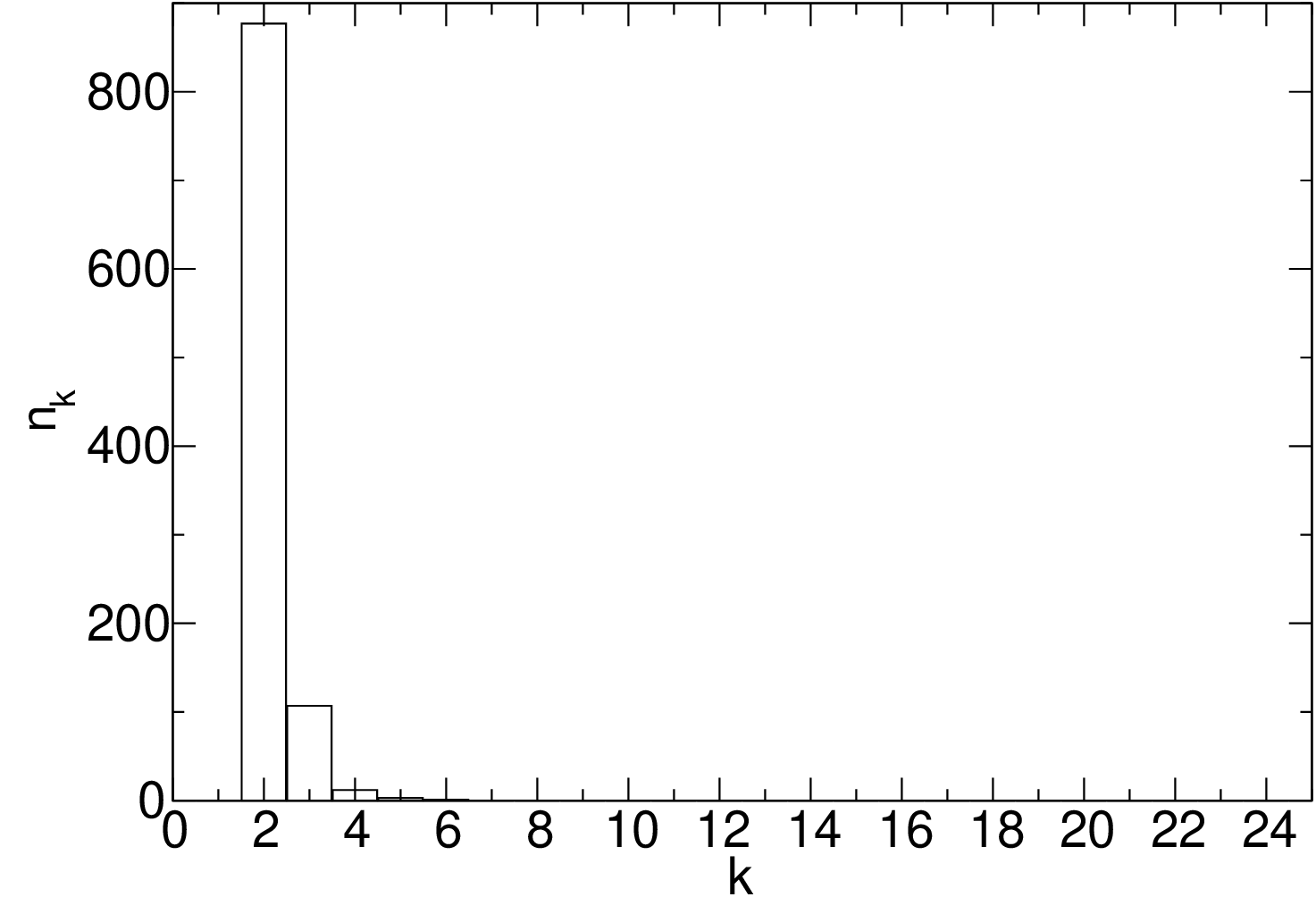}
    }
  \end{figure}
\end{center}

\newpage
\begin{center}
  \begin{figure}
    \resizebox{0.75\columnwidth}{!}{%
      \includegraphics{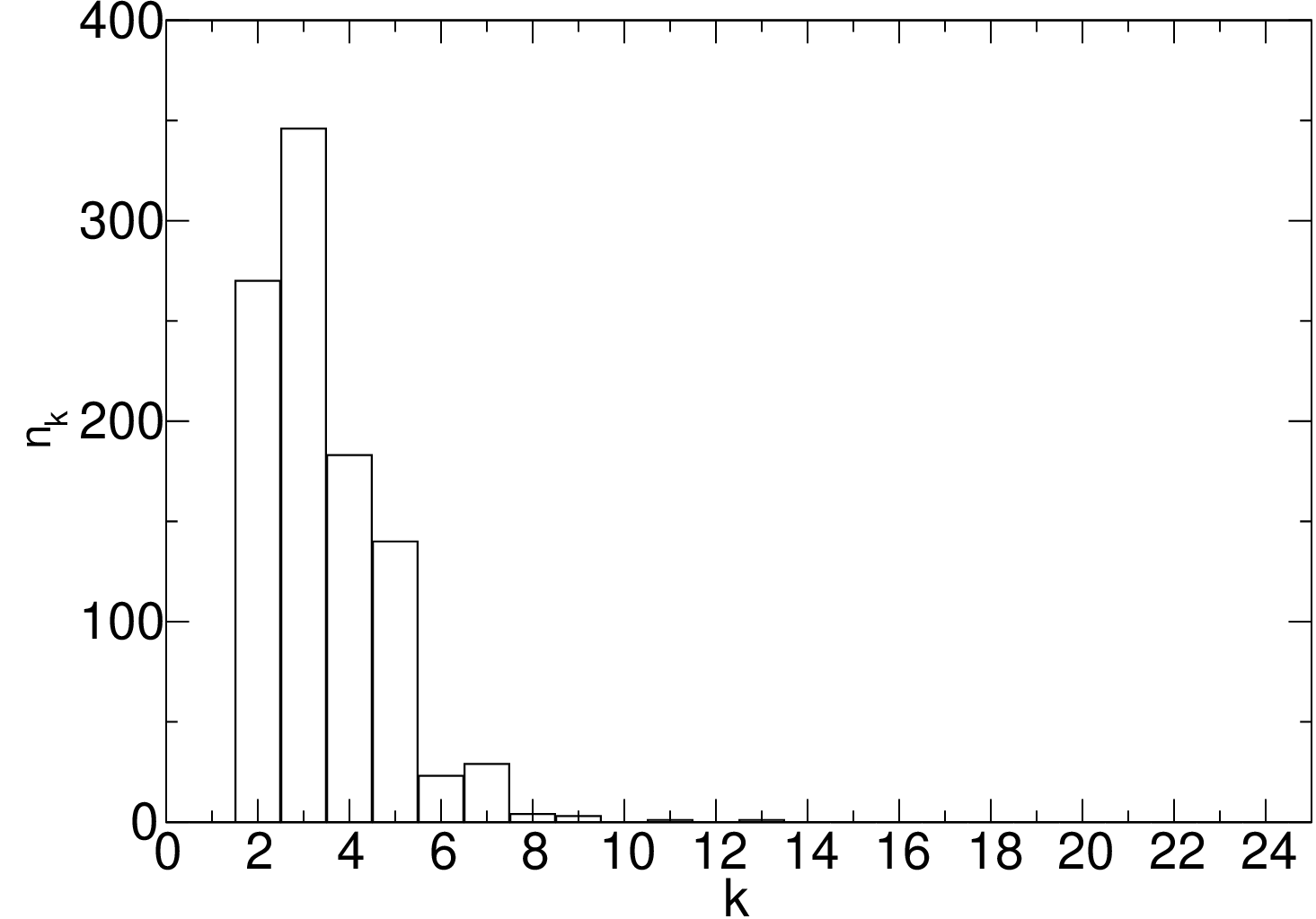}
    }
  \end{figure}
\end{center}

\newpage
\begin{center}
  \begin{figure}
    \resizebox{0.75\columnwidth}{!}{%
      \includegraphics{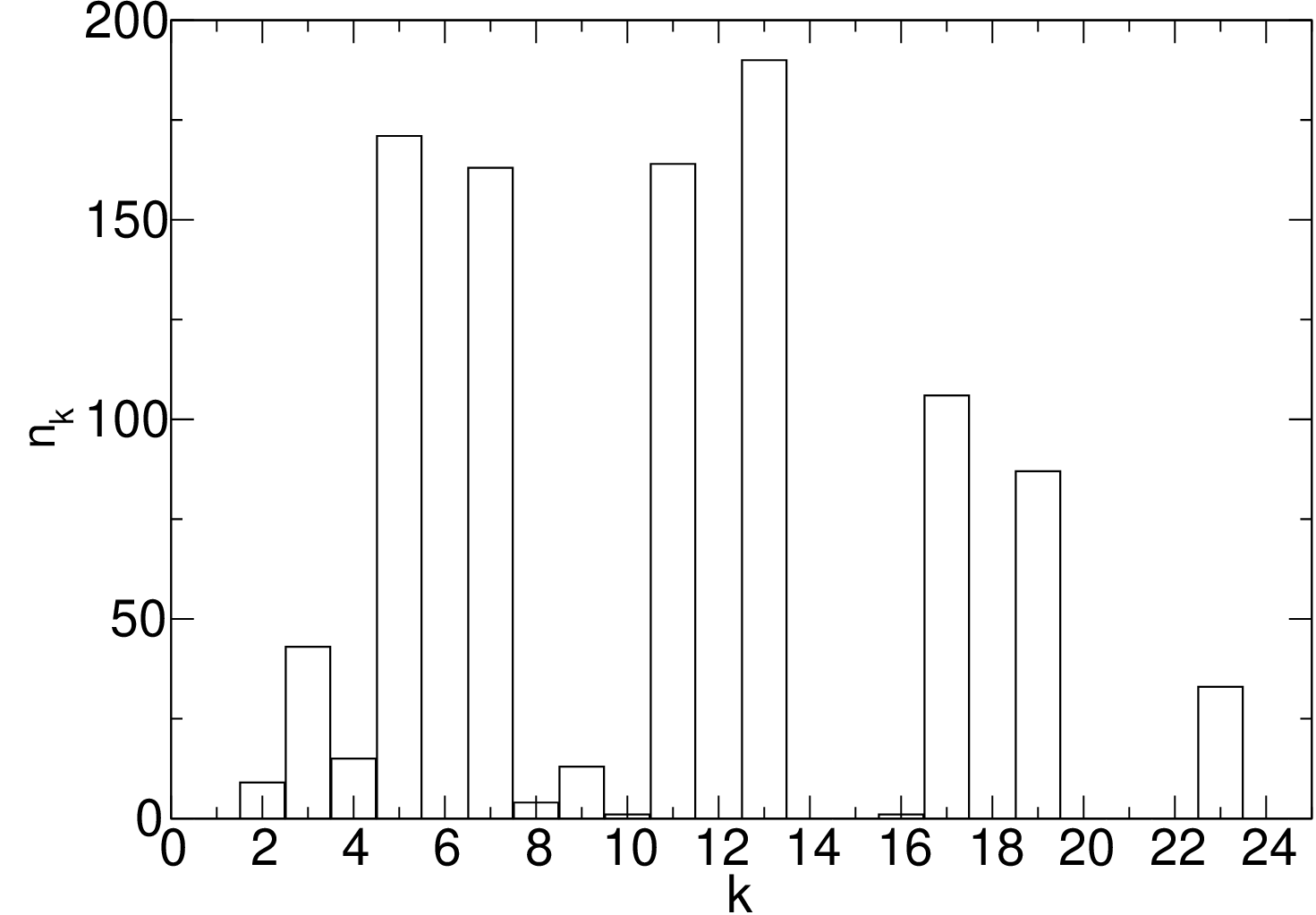}
    }
  \end{figure}
\end{center}

\newpage
\begin{center}
  \begin{figure}
    \resizebox{0.75\columnwidth}{!}{%
      \includegraphics{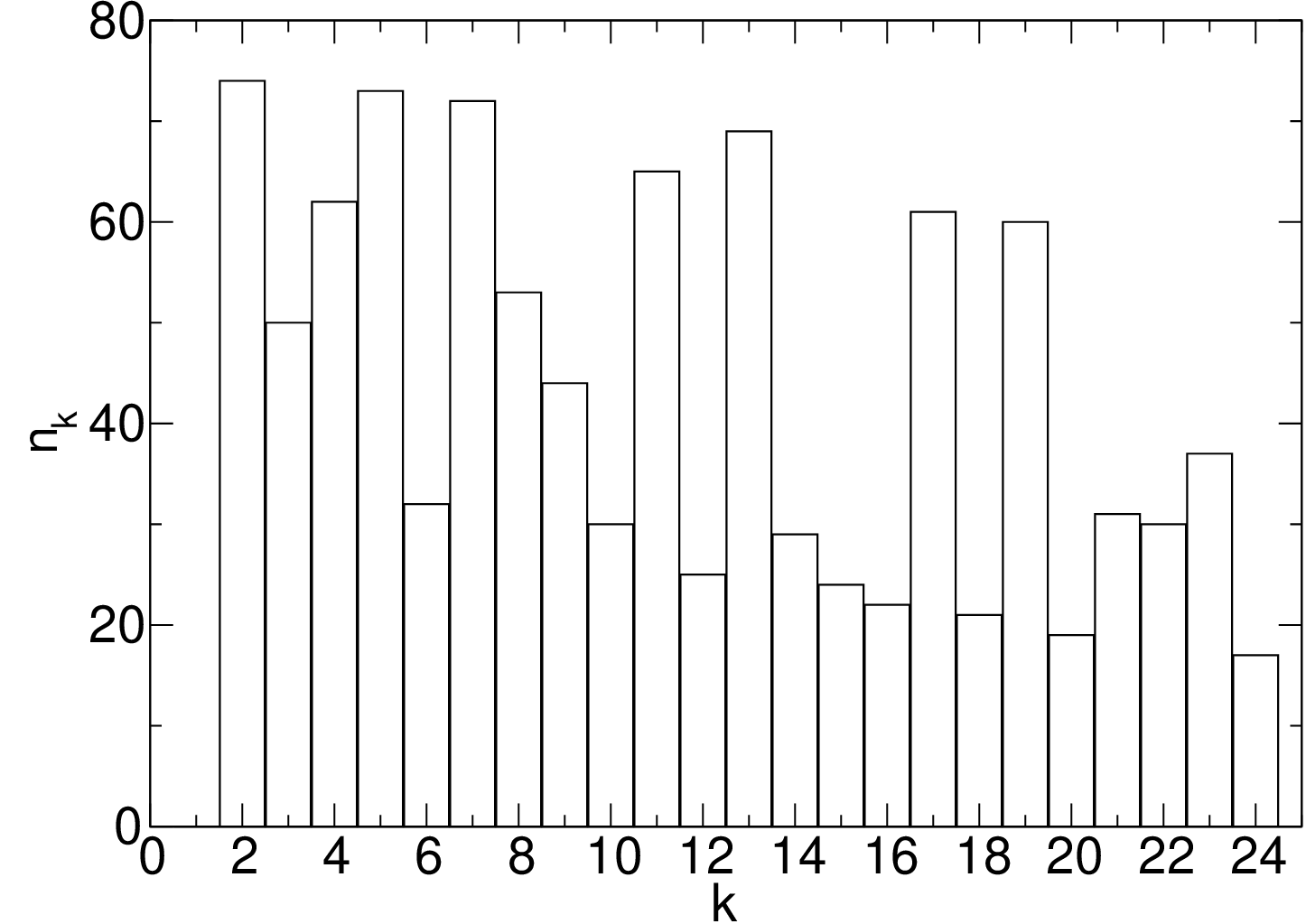}
    }
  \end{figure}
\end{center}

\newpage
\begin{center}
  \begin{figure}
    \resizebox{0.75\columnwidth}{!}{%
      \includegraphics{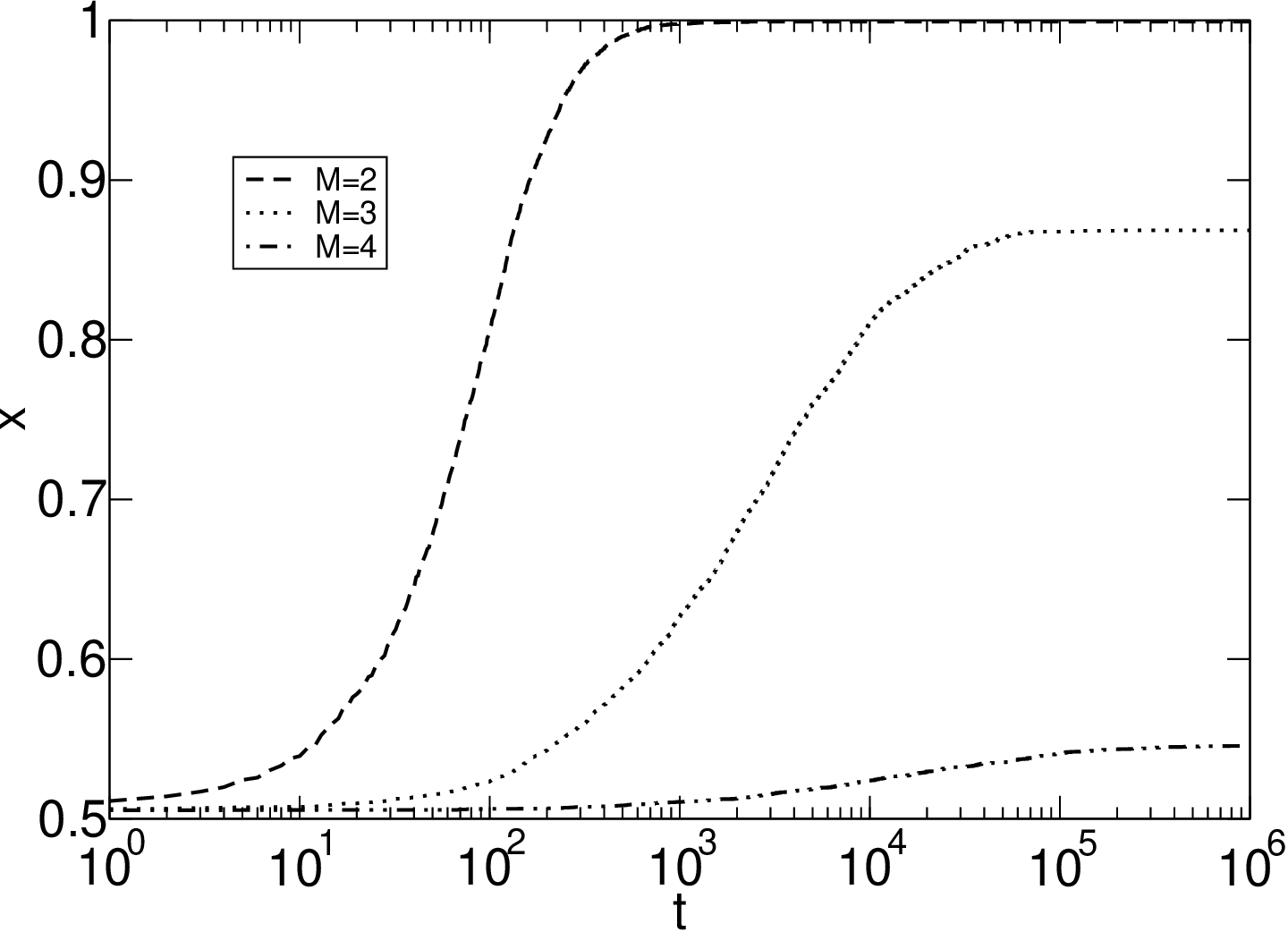}
    }
  \end{figure}
\end{center}

\end{document}